\documentclass[conference]{IEEEtran}
\IEEEoverridecommandlockouts
\usepackage{cite}
\usepackage{amsmath,amssymb,amsfonts}
\usepackage{algorithmic}
\usepackage{graphicx}
\usepackage{textcomp}
\usepackage{xcolor}
\usepackage{amsfonts}
\usepackage{booktabs}
\usepackage{float}
\usepackage{graphicx}
\usepackage{grffile}
\usepackage[algo2e]{algorithm2e}
\usepackage{longtable}
\usepackage{wrapfig}
\usepackage{gensymb}
\usepackage{textcomp}
\usepackage{amsmath}
\usepackage{algorithm}
\usepackage{verbatim}
\usepackage{graphicx}
\usepackage{caption}
\usepackage{listings}
\usepackage{array}
\usepackage{subcaption}
\usepackage{booktabs}
\usepackage{amsmath}
\usepackage{enumitem}
\usepackage{dblfloatfix}
\usepackage{lipsum}
\def\arraystretch{1.25}
\usepackage[utf8]{inputenc}

\begin{document}
\title{Learner's Dilemma: IoT Devices Training Strategies in Collaborative Deep Learning}
\makeatletter
\newcommand{\linebreakand}{%
  \end{@IEEEauthorhalign}
  \hfill\mbox{}\par
  \mbox{}\hfill\begin{@IEEEauthorhalign}
}
\makeatother

\author{\IEEEauthorblockN{Deepti Gupta\IEEEauthorrefmark{1}, Olumide Kayode\IEEEauthorrefmark{2}, Smriti Bhatt\IEEEauthorrefmark{3}, Maanak Gupta\IEEEauthorrefmark{4}, Ali Saman Tosun\IEEEauthorrefmark{5} }
\IEEEauthorblockA{\IEEEauthorrefmark{1}\IEEEauthorrefmark{2}\IEEEauthorrefmark{5}Dept. of Computer Science,
University of Texas at San Antonio,
San Antonio, Texas 78249, USA \\\IEEEauthorrefmark{3}Dept. of Computing and Cyber Security, Texas A \& M University-San Antonio,
San Antonio, Texas 78224, USA\\\IEEEauthorrefmark{4}{Dept. of Computer Science},
{Tennessee Technological University},
Cookeville, Tennessee 38505, USA \\}
\IEEEauthorrefmark{1}deepti.gupta@my.utsa.edu, \IEEEauthorrefmark{2}olumide.kayode@utsa.edu, \IEEEauthorrefmark{3}sbhatt@tamusa.edu, \IEEEauthorrefmark{4}mgupta@tntech.edu, \IEEEauthorrefmark{5}tosun@cs.utsa.edu}

\maketitle

\begin{abstract}
With the growth of Internet of Things (IoT) and mobile edge computing, billions of smart devices are interconnected to develop applications used in various domains including smart homes, healthcare and smart manufacturing. Deep learning has been extensively utilized in various IoT applications which require huge amount of data for model training. Due to privacy requirements, smart IoT devices do not release data to a remote third party for their use. To overcome this problem, collaborative approach to deep learning, also known as Collaborative Deep Learning (CDL) has been largely employed in data-driven applications. This approach enables multiple edge IoT devices to train their models locally on mobile edge devices. In this paper, we address IoT device training problem in CDL by analyzing the behavior of mobile edge devices using a game-theoretic model, where each mobile edge device aims at maximizing the accuracy of its local model at the same time limiting the overhead of participating in CDL.
%(at a minimum cost of participating in CDL) 
We analyze the Nash Equilibrium in an \textit{N}-player static game model. We further present a novel cluster-based fair strategy to approximately solve the CDL game to enforce mobile edge devices for cooperation. Our experimental results and evaluation analysis in a real-world smart home deployment show that 80\% mobile edge devices are ready to cooperate in CDL, while 20\% of them do not train their local models collaboratively. 
%do not participate in CDL.
\end{abstract}


\section{Introduction}
\label{sec:introduction}
In recent years, Internet of Things (IoT) has grown rapidly and billions of smart devices are expected to be added over next few years. These devices generate a tremendous amount of data, from health information \cite{celik2018soteria} to social networking \cite{zeng2017end}. Deep learning models use this data for training and enhancing intelligence of various data driven IoT applications. Most of the IoT devices connect to a central cloud platform to use cloud services. These cloud services are crucial for storage of the datasets and model learning. However, use of cloud services requires additional latency in real time applications. To overcome this issue, edge devices are used for local data training which also safeguards privacy of personal data. Unlike constrained IoT devices, such devices have the capability to support Machine Learning (ML) models and have been used in various applications. For example, video doorbell performs training on its local datasets, and identifies person at the door. \par

Deep learning models are often associated with the size of training dataset. Under a reasonable learning mechanism, more training data will enhance the accuracy and performance of a trained model. However, in the era of big data, data is often distributed and cannot be brought together due to personal privacy constraints. Collaborative Deep Learning (CDL) allows multiple IoT devices to train their models, without revealing associated personal data. CDL offers an attractive trade-off between privacy and utility of data sets. Recent research \cite{jiang2019lightweight,chen2019communication} have discussed the privacy issues of local training devices and %
the impact of communication latency between IoT edge devices and Parameter Server (PS). However, the strategic behavior of the rational local training devices have not been discussed in previous research, i.e., the authors have assumed that all IoT devices are altruistic. Altruistic devices are ones which always follow a suggested protocol (what all devices have decided to follow initially) regardless whether they are benefiting or losing by following this protocol. However, devices are not altruistic in real life, they are rational. Rational devices are the ones which will deviate from suggested protocol if they think that they will be benefited more by following a different protocol. In our proposed system model,  we assume that all the mobile edge devices are rational.

A mobile edge device, which has low quality data, always wants to be a part of CDL to increase accuracy of its local model. Other mobile edge devices, who have high quality data,  do not want to collaborate with low quality data holder mobile edge device. Therefore, there is a dilemma for mobile edge devices to participate or not in CDL. In this paper, we address this research problem of learner's dilemma by proposing a general system model, a CDL game model, and a novel cluster-based fair strategy which enables each participant to cooperate in CDL based on the clusters formed to achieve overall benefit to itself in training the local ML model. We also evaluate our CDL game model and novel cluster-based strategy in smart home deployment using ARAS dataset\cite{SmartHome}. The main contributions of this paper are as follows. 
\begin{enumerate}
    \item We identify the problem of unfair cooperation of participants in CDL. A local training device, which has low quality data builds its learning model to take advantage from other device, which has high quality data.
    \item We address this research problem by analyzing the behavior of mobile edge devices using a game-theoretic model, where each device aims at maximizing the accuracy of its local model with minimal cost of participating in CDL.
    \item We introduce a system model for CDL and propose a solution of above defined problem.
    \item We also implement a cluster-based fair algorithm on ARAS dataset \cite{SmartHome}, and the results reflect that proposed solution elicit cooperation in CDL.
\end{enumerate}

\par
The rest of paper is organized as follows. Section \ref{sec:related}
presents relevant work and related background. % of privacy issues in deep learning models, privacy preserving deep learning, and game theoretic models. 
System model along with rational assumption is discussed in Section \ref{sec:system-model}. Game model and game analysis are explained in Section \ref{sec:The Collaborative Deep Learning Game} and Section \ref{sec:game-analysis} respectively. Section \ref{sec:num-anal} presents implementation of proposed system model along with results. Section \ref{sec:conc} concludes the paper with future research directions. 

\section{Related Work}
\label{sec:related}
In this section, we describe related work on information leakage on deep learning models, privacy-preserving deep learning and game models.
\subsection{Information leakage on Deep Learning Models} 
Information leakage of individuals' private data has become a well known problem for deep learning models. Data masking techniques, such as pseudonymize and anonymize are used to prevent this problem. In pseudonymize, data can be traced back into its original state, whereas it becomes impossible to return data into its original state in anonymize. However, indirect re-identification could be possible in anonymize. For example, Netflix released a hundred million anonymized film ratings which was matched with the other dataset Internet Movie Database (IMDb). \par 

Cloud platforms such as Google and Amazon offer various services "AI Deep Learning". Any customer can upload a dataset to use the service and pay to build a prediction model, which works as black-box API. The membership inference attack on black-box API is discussed in \cite{shokri2017membership,yeom2018privacy}. An attacker asks queries to target the model and receives the model’s prediction. Rahman et al. \cite{rahman2018membership} show that differentially private deep model can also fail against membership inference attack. A novel white-box membership inference attack was proposed by Nasr et al. \cite{nasr2018comprehensive}, against deep learning algorithms to measure their training datasets membership leakage. Melis et al. \cite{melis2019exploiting} demonstrate that the updated parameter leaks information of participants, thus develops passive and active inference attacks to exploit this leakage.

\subsection{Privacy-Preserving Deep Learning}
Each participant has its own sensitive datasets, which needs to be protected the dataset from information leakage. Various privacy mechanisms, such as Secure Multi-party Communication (SMC) \cite{kerschbaum2009practical}, Homomorphic Encryption (HE) \cite{rivest1978data}, and Differential Privacy (DP) \cite{dwork2014algorithmic} have been proposed to protect the datasets in CDL. SMC helps to protect intermediate steps of the computation. Mohassel et al. \cite{mohassel2017secureml} adopt a two-server model for privacy-preserving training, used by previous work on privacy-preserving deep learning via SMC \cite{nikolaenko2013privacy1}.

However, Aono et al. \cite{aono2018privacy} pointed out that the local data may be actually leaked to an honest-but-curious server. Using additively HE techniques fix several problems and also have some drawbacks. To obscure an individual's identity, DP adds mathematical noise to a small sample of the individual's usage pattern. Prior work \cite{abadi2016deep,jiang2019lightweight,shokri2015privacy,weng2018deepchain} use differential privacy on privacy-preserving CDL system to protect privacy of training data. However, Hitaj et al. \cite{hitaj2017deep} pointed out that privacy preserving deep learning approach is failed to protect data privacy and demonstrated that a malicious participant can learn personal information of other participant through Generative Adversarial Network (GAN) learning.\par 

The most dominant technique to optimize the loss function is Stochastic Gradient Descent (SGD). SGD is a method to find the optimal parameter configuration for a ML algorithm. SGD is applied in various privacy-preserving deep learning models \cite{abadi2016deep,melis2019exploiting, mohassel2017secureml,nasr2018comprehensive}. PS receives the gradients from mobile edge devices by using different approaches like round robin, random order \cite{shokri2015privacy}, cosine distance \cite{chen2018machine}, time based \cite{weng2018deepchain}. The server aggregates the received parameters using Federated Averaging algorithm \cite{mcmahan2016communication}, and weighted aggregation strategy \cite{chen2018machine}. Federated averaging algorithm introduced for model averaging combines local SGD on each client with a server. It is robust for unbalanced and non-IID data distributions, and reduce rounds of communication needed to train a Deep Learning (DL) model.
\subsection{Game Models}
In prior academic research, game theory has been applied into data privacy game to analyze privacy and accuracy. Pejo et al. \cite{pejo2018price} defined two player game, in which one is privacy concerned and other not. Esposito  et al. \cite{esposito2018securing} proposed a game model to analyze the interaction between a provider (global ML model) and a requester (local ML model) within a CDL model. In this literature, there have been various game models about privacy-accuracy trade-off and energy-efficient solution. 
However, to the best of our knowledge, there is no prior work to utilize game theory to analyze mobile edge devices' rational behavior in a selfish environment. Therefore, we construct a game model for rational mobile edge devices in CDL and analyze the game.
\begin{figure}[t]

\centering
\includegraphics[width=.5\textwidth, height=.23\textheight]{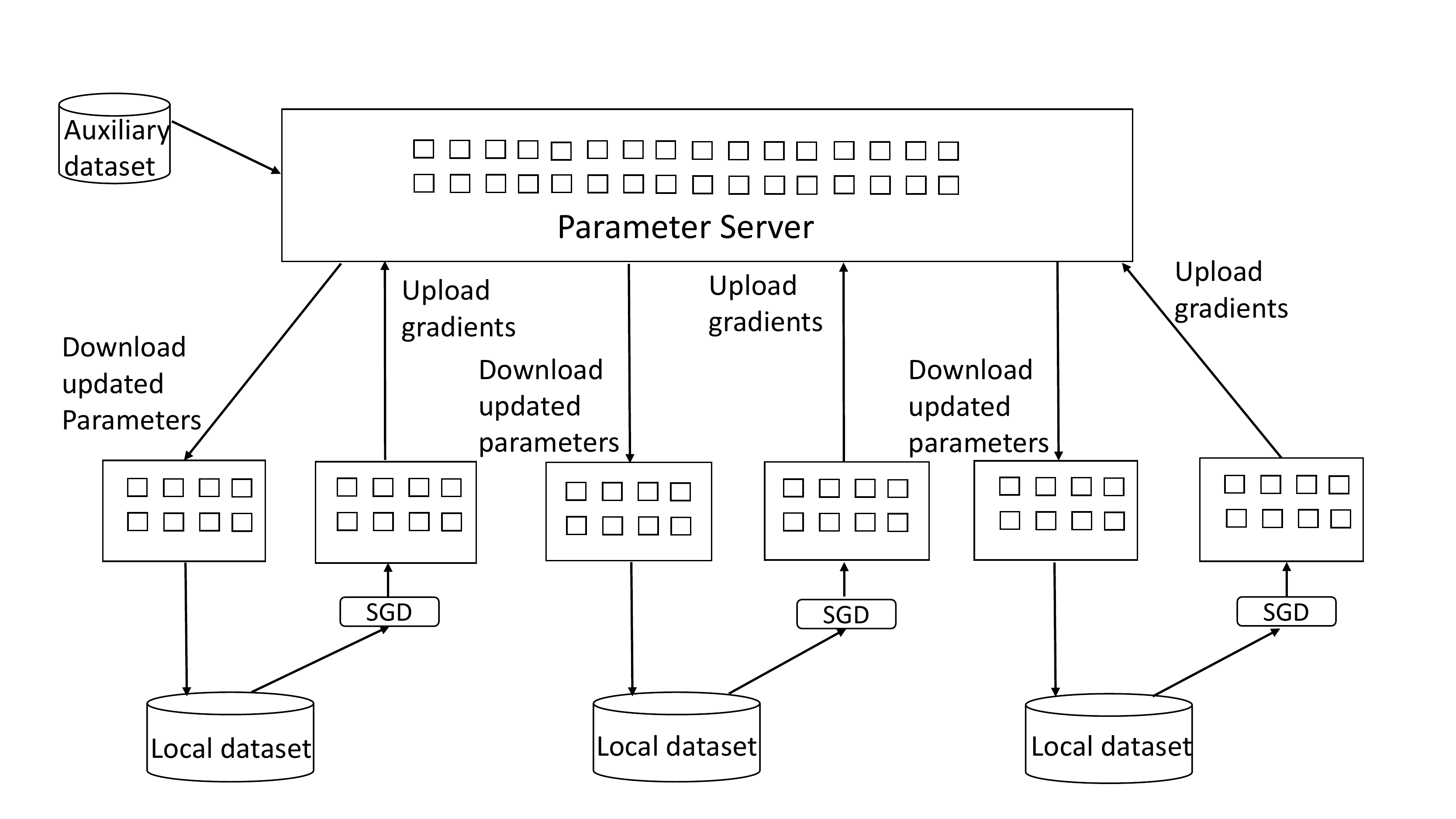}
\centering
\caption{A Collaborative Deep Learning System Model}
\end{figure}

\section{System Model}
\label{sec:system-model}
We first generically outline details of the CDL model, in which all edge computing devices, such as mobile phones and IoT devices are assumed to be altruistic. Then, we clarify the rationality assumptions of mobile edge devices in CDL model.

\subsection{Collaborative Deep Learning Model}
Figure 1 illustrates the main components of the system model. Consider there are \textit{N} number of mobile edge devices, and each mobile edge device is connected with multiple IoT devices. These IoT devices generate huge amount of data, which is used for training to build ML model. These devices train their data to build local models in a collaborative manner without compromising data privacy, which is beneficial for mobile edge devices and IoT devices. In our model, we assume that each mobile edge device maintains a local vector of neural-network parameters, $w^i$. The PS maintains a separate parameter vector $w^{global}$. Each edge device can initialize parameters (weights) $w^i$  where \textit{i}=1,2,3,..\textit{N} randomly or by downloading their latest values from PS. Each edge device trains a local model and optimize the loss value using SGD. Here, one weight sample is selected at random in each optimization step. This process continues until SGD converges to a local optimum. Let E be the error or loss value, i.e., the difference between the true value of the objective function and the computed output of the network, it can be based on $L^2$ norm or cross entropy. The back-propagation algorithm computes the partial derivate of E with respect to each parameter in $w^k$ and updates the parameter so as to reduce its gradient. We refer to one full iteration over all available input data as an epoch. All mobile edge devices train their local models simultaneously through PS.

\begin{algorithm}[H]
\SetAlgoLined
\caption{Pseudo Code for Mobile Edge Device i}
\begin{algorithmic}[1]
\STATE{Define initial parameters $w^i$, learning rate $\alpha$ and number of local epochs H.}
\STATE{Repeat all the steps until minimum error is obtained:}
\STATE{Download parameters $w^{global}$ from PS to learn a common learning objective.}
\STATE{LocalTraining (i,$w^i$): Train local data on each device to build own ML model}
\STATE{Split local dataset $D_i$ to minibatches of size K which are included into the set $K_i$.}
\FOR{for each local epoch j from 1 to H}
    \FOR {for each k $\in$ $K_i$}
\STATE\begin{equation}
\Delta w^{i} = \Delta w^{i}  - \alpha\frac{\partial E_{i}}{\partial w^{i}}
\end{equation}
\ENDFOR
\ENDFOR
\STATE{Each participant uploads local gradients $\Delta$$w^i$ to PS}
\STATE{Each participant downloads updated parameters $w^{global}$ and loss value $\tau_i$ of each participant.}
    
\end{algorithmic}   
\end{algorithm}

There is no need for any coordination among all local training devices. They can influence each other's training indirectly, via PS. PS receives local gradients $\Delta{w^i}$ from each edge devices and aggregates them with global parameter \textit{w\textsuperscript{global}}.
After updating this global parameter, each participant  downloads \textit{w\textsuperscript{global}} parameter from PS and starts training based on global parameter. There are various scenarios to exchange the parameters from PS to mobile edge device. In this model, PS exchanges the parameters asynchronously, i.e. PS does not wait for all local gradients from all edge devices. When a participant trains his local model, others may update their parameters through PS. This process continues until the model achieves the desired output.

\begin{algorithm}[H]
\SetAlgoLined
\caption{Pseudo Code of Parameter Server}
\begin{algorithmic}[1]
\STATE{Set initial global parameters $w\textsuperscript{global}$}
\STATE{PS runs these local gradients $\Delta{w^i}$ on auxiliary dataset and calculate loss values $\tau_i$ of each participant} 
\STATE{PS also aggregates these local gradients $\Delta{w^i}$ asynchronously}
        \begin{equation}
        w\textsuperscript{global} = w\textsuperscript{global} + \Delta w^{i}
        \end{equation}
 \end{algorithmic}  
\end{algorithm}

\subsection{Mobile Edge Device costs}
We now characterize the costs (computation and communication costs) borne by mobile edge devices and IoT devices to their participation in CDL system. It should be noted that our goal is not to arrive at a precise quantification of these costs, rather to characterize them such that they could be used to analyze the strategic behavior of the devices while participating in CDL. The CDL system is basically grouped into two phases: (1) Training phase, and (2) Participating phase. During the training phase, each device builds a local model and initialize their weights to train the network. During training, each participant calculates its local gradients to upload on PS. PS aggregates all the local gradients and sends back to each device. The updated parameters are downloaded by each device and training process continues until loss value becomes negligible.  \par
Thus, we can characterize the total cost for a mobile edge device to participate in an epoch to build ML model based on the cost for executing the above two phases. For the training phase, a mobile edge device bears a cost \textit{c\textsuperscript{plocal}}, which is computation cost to build a local ML model. Another computation cost is \textit{c\textsuperscript{pglobal}} for training a local model using updated global parameters. Accordingly, for executing the participation phase, a mobile edge devices bears another costs \textit{c\textsuperscript{m'}} and \textit{c\textsuperscript{m}}. The cost \textit{c\textsuperscript{m}} is communication cost, where a mobile edge device uploads its parameters to PS, the cost \textit{c\textsuperscript{m'}} is also communication cost, where a mobile edge device $i$ downloads the updated parameters from PS. 
The average per mobile edge devices cost $c_{i}^{t}$ for participation in each epoch of collaborative deep learning system can be characterized as
\begin{equation}
c_i^t = c^{plocal} + c^{m} + c^{m'} + c^{pglobal}
\end{equation}

One point that needs further clarification is why a participant may choose not to spend these costs \textit{c\textsuperscript{m}}, \textit{c\textsuperscript{m'}}, and \textit{c\textsuperscript{pglobal}}. Our rationality assumption provides this clarification. 

\subsection{Rationality Assumption}
Prior research in distributed DL \cite{chen2018machine} have assumed a byzantine adversary where mobile edge devices or IoT devices controlled by adversary can be arbitrarily malicious, i.e. malicious participant could arbitrarily deviate from suggested protocol in CDL or could arbitrarily drop communication between mobile edge device and PS. However, here we assume that mobile edge devices and IoT devices are honest but they are selfish. In this setup, the notion of rationality means that a rational device choose to participate or not to maximize its profit in CDL.

\begin{table}[]
\scriptsize
\def\arraystretch{1.15}
\caption{List of Symbols.}

\begin{tabular}{c|l} \toprule
\textbf{Symbol}  &    \textbf{Definition}  \\
\bottomrule
$N$				 & Number of mobile edge devices\\
$n$& Total number of IoT devices\\
$K$ & Batch size \\
$H$ & Numbers of local epoch\\
$D\textsubscript{i}$ & Generated data from IoT device i\\

$\Delta w\textsubscript{i}$ & Local gradient of participant i \\

$w\textsuperscript{global}$	& Global parameter\\

$\alpha$		 & Learning rate \\

$M$				 & Privacy mechanism \\

$\theta\textsubscript{i}$ & Loss value of participant i, train individually\\

$\phi\textsubscript{i}$ & Loss value of participant i,train collaboratively\\

$\tau_i$ & Loss value of participant i, train individually on auxiliary dataset \\

$B$ & coefficient \\

$c\textsuperscript{plocal}$  & Computation cost to build a local model \\

$c\textsuperscript{pglobal}$  & Computation cost to build a global model \\

$c\textsuperscript{m}$ 	& Communication cost to upload the parameters to PS\\

$c\textsuperscript{m'}$ 	& Communication cost to download the parameters from PS \\
$c_i\textsuperscript{t}$ 	& Total cost for build a ML model \\

$C_i$					& Number of cooperative participants\\

$N-C_i$					& Number of defective participants\\

\bottomrule
\end{tabular}

\label{tab:Symbols}

\vspace{2mm}

\end{table}

\section{The Collaborative Deep Learning Game}\label{sec:The Collaborative Deep Learning Game}

We present a game model of CDL system with multiple mobile edge devices in a honest but selfish environment. We introduce a game model with $N$-players that we refer to as the collaborative learning game G. In this game, the edge devices send their local gradients to PS to learn a common objective without compromising the privacy of data. PS aggregates the gradients and creates a global model. This updated global model is downloaded by all participating edge devices, where exists a social dilemma for all defection behavior.
  
\subsection{Game Model}
Game theory allows for modeling situations of conflict and for predicting the behavior of participants when they interact with each other. In our CDL game G, mobile edge devices who are connected with multiple IoT devices are participants, they interact with PS simultaneously without having any knowledge about each other. The Game G is a static game, because all participants must choose their strategy simultaneously. The Game G is a tuple $(P,S,U)$, where $P$ is the set of players, $S$ is the set of strategies and $U$ is the set of payoff values.

\begin{itemize}
\item \textbf{Players} ($P$): The set of players $P=\sum_{i=1}^{N} P_{i}$ corresponds to the set of mobile edge devices who received a common objective from PS to build its own local model in CDL game G. 

\item \textbf{Strategy} ($S$): Each participant $P_i$ can choose between two actions $s_i$ (i) Cooperative ($CP$) or (ii) Defective ($DF$). Hence the set of strategies in this game is $S$ = \{$CP$,$DF$\}. Strategy of mobile edge devices $P_i$ determines whether $P_i$ participates in CDL. In particular, if a participant $P_i$ plays $CP$ strategy, i.e., it will send its local gradients to PS and downloads updated parameters from PS to update its local model. Here, the participant pays total cost. In contrast, if a participant $P_i$ neither uploads its local gradients to PS nor downloads the updated global parameters from PS, i.e. the mobile edge devices $P_i$ plays $DF$ strategy. Thus, participants saves its communication costs $c^m$, $c^m{}^{'}$ and global computation cost $c^{pglobal}$. Here, this participant is not part of CDL and trains its local model individually.

\item \textbf{Payoff} ($U$): At a high level, the players’ goal in CDL game G is to maximize their utility, which is a function of the loss value and its costs. In this work, we do not consider the adversarial aspect of players; hence, the gain includes only the accuracy improvements on the model for a particular player as a benefit while the costs are used for training the models and communication between participant and PS. 
\end{itemize}

Here, the benefit and the cost are not on the same scale as the first depends on the loss value while the latter on cost. To make them comparable, we introduce a coefficient: the benefit is multiplied with B.

Now, we compute the payoff of mobile edge devices $P_i$ in this game. If we assume that the participant $P_i$ is cooperative, i.e. $P_i$ $\in$ $CP$. Similarly, if $P_i$ is defective, i.e. $P_i$ $\in$ $DF$, and these payoff can be defined as follows.

\begin{equation}
u_i(CP) = B(\frac{1}{\phi_i}) - (c^{plocal} + c^m + c^{m'} + c^{pglobal})
\end{equation}

\begin{equation}
u_i(DF) = B(\frac{1}{\theta_i}) - (c^{plocal})  
\end{equation}

Based on the above calculated utilities, we analyze the game G as discussed in the next section.

\section{Game Analysis}\label{sec:game-analysis}
In order to get an insight into strategic behavior of participants, we apply the most fundamental game-theoretic concept, known as Nash Equilibrium, introduced by John Nash \cite{nash1951non}.

\textbf{Definition 1.} A Nash Equilibrium is a concept of game theory where none of the players can unilaterally change their strategy to increase their payoff. 

In other words, if in a non-cooperative game all strategies are mutual best responses to each other, then no player has any motivation to deviate unilaterally from the given strategy one Nash Equilibrium strategy profile. For example, in  any  prisoners’  dilemma  game,  there  is  always  a  cooperative strategy and a defecting  strategy. If both players use cooperative strategy, then it yields the best outcome for the players. If the players do not cooperate with one another, then they choose defecting strategy in the hope of attaining individual gain at the rival’s expense. In prisoners’ dilemma defecting  strategy strictly dominates the cooperation strategy. Hence, the only Nash Equilibrium in prisoners’ dilemma, is a mutual defection. 

Based on the cost and benefit of mobile edge devices to learn a neural-network model, we build a one-shot CDL game model G. In the following theorems, we show that the game G is a public good game.

\textbf{Theorem 1.} \textit{ In CDL game G, if a participant builds its local ML model, then G reduces to a public good game.} 

\textit{Proof.} Let us consider all \textit{N} participants follow defective-$DF$ strategy where all participants neither send their local gradients to PS nor download updated parameters from PS, i.e., no communication between mobile edge device and PS. So, participants do not pay any communication costs $c^m$, $c^{m'}$, and global computation cost $c^{pglobal}$. Now each participant $P_i$ trains local data sets $D_i$ to build its ML model individually and minimizes its loss value $\theta_i$. None of participants cannot change his strategy profile unilaterally. Let us assume if a participant deviates from defect-$DF$ strategy to cooperate-$CP$ strategy unilaterally, then participant will pay all these costs ($c^m$ + $c^{m'}$ + $c^{pglobal}$). The payoff of cooperate-$CP$ strategy is less than defect-$DF$ strategy, so All-$DF$ is a Nash equilibrium profile and G is a public good game. %\vspace{2mm} 

Theorem 2 further shows we can never enforce an all cooperate-$CP$ strategy in game G, and therefore, we could not establish a Nash Equilibrium.

\textbf{Theorem 2.}\textit{ In CDL game G, if a participant builds its local ML model, then we cannot establish All-Cooperation strategy profile as a Nash Equilibrium.} \vspace{2mm}

\textit{Proof.} We first assume that all \textit{N} participants have already cooperated in collaborative learning (i.e., all cooperate-$CP$ strategy profile) and payed communication costs as well as global computation cost. We can compute the payoff of each participant $P_i$ by Equation (4). Hence, if a participant deviates from the cooperation and play defection unilaterally, its payoff would be equal to Equation (5), which is always greater than cooperative payoffs at Equation (4). Hence, each participant has incentive to deviate unilaterally and increases its payoff. Then, the All cooperate-$CP$ strategy profile is never a Nash Equilibrium.

 \vspace{2mm} 

\subsection{Cluster-Based Representation}
Each participant has loss values of all other participants, which is calculated on auxiliary dataset. Before the start of the game, each participant has to choose his strategy to play the collaborative game G. However, in the beginning of the game, the participant is not sure about his strategy, which will depend on other participant's strategy.
%based on their loss values. 
Therefore, all the participants are in dilemma to choose a strategy between $CP$ and $DF$. We solve this problem by proposing the cluster-based fair strategy algorithm. K-means clustering is an unsupervised ML technique, whose purpose is to segment a data set into K clusters. Each participant applies k-means cluster algorithm on all loss values (one-dimensional data). 

\begin{algorithm}[H]
\SetAlgoLined
\caption{Cluster-Based Fair Strategy}
\begin{algorithmic}[1]
\STATE{Apply k-means clustering algorithm on loss values of each participant i.}
\STATE{Make the clusters}
\IF{participant i belongs a cluster with at least one other participant j}
\STATE $P_i$ , $P_j$ $\in$ $CP$
\ELSE
\STATE $P_i$ $\in$ $DF$
\ENDIF
\end{algorithmic}
\end{algorithm}

\begin{figure}[t]

\centering
\includegraphics[width=.60\textwidth, height=.30\textheight]{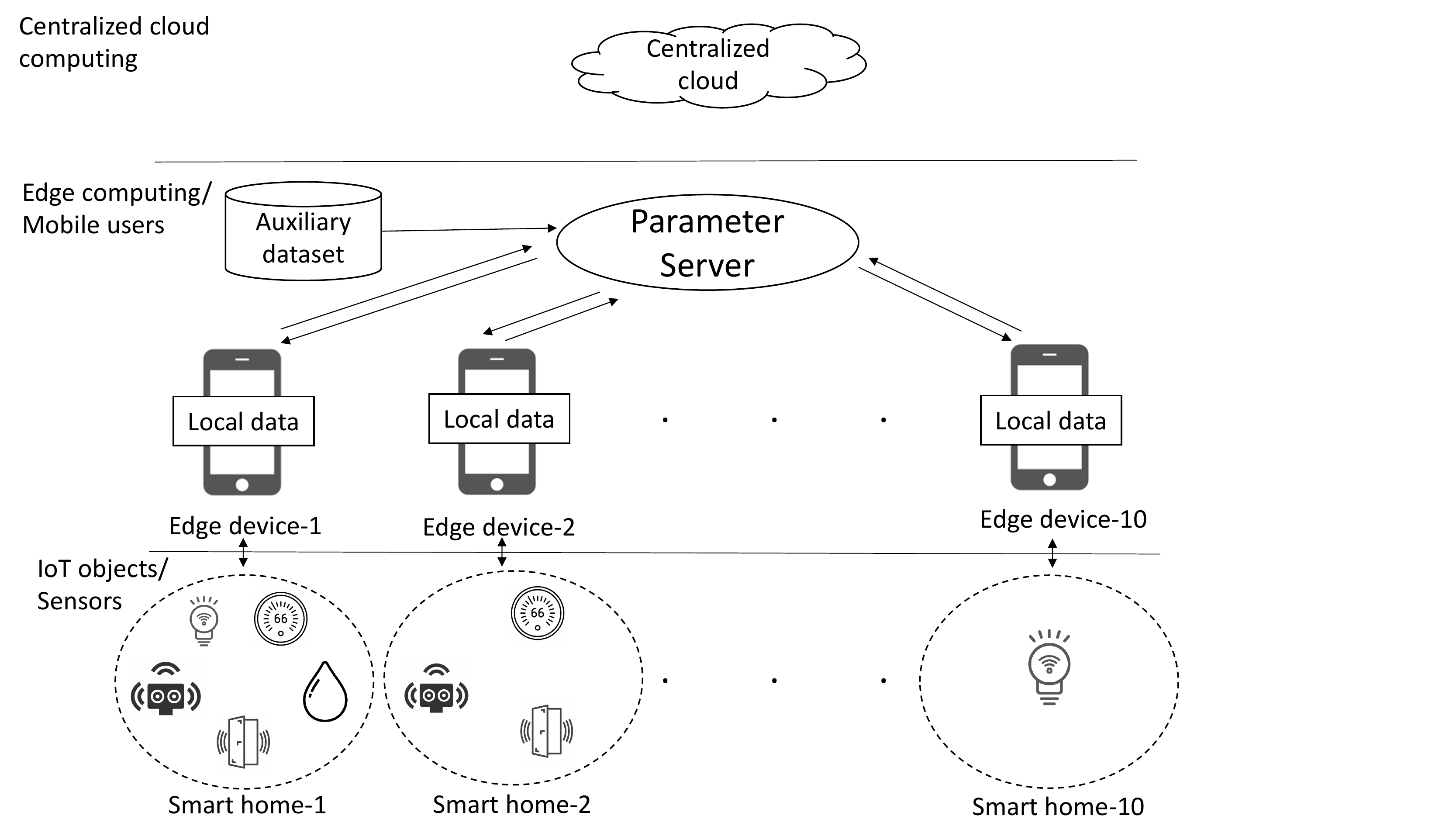}
\centering
\caption{Experimental Setup of Proposed Game Model}
\end{figure}

\section{Numerical Analysis}\label{sec:num-anal}
To evaluate our proposed cluster-based fair strategy, we apply this novel strategy on smart home datasets.% with numerous IoT devices. 

\subsection{Datasets}
We use ARAS datasets to build smart home interaction model, since this data is available publicly \cite{SmartHome}. ARAS dataset is real world dataset for activity recognition with ambient sensing. The living residents did not follow any specific rules to live in smart homes. This dataset contains two real smart home data with multiple residents for one month. It contains 3000 daily life activities captured by 26 million sensor readings in smart homes. This dataset also has ground truth labels for activities, which enables to develop a new sophisticated ML smart home interaction model. 

\subsection{Experimental Setup}
We simulate the proposed system model with \textit{N} number of mobile edge devices associated with smart homes. IoT devices are connected with one mobile edge device in each smart home. We partitioned ARAS dataset unevenly into 10 participants. For unbalanced datasets setting, the data is sorted by class and divided into two cases: (a) low quality dataset, where the participant receives data partition from a single class, and (b) high quality, where participant receives data partition from 27 classes. Figure 2 shows unbalanced partitioning of the dataset, smart home-1 generates high quality data (multiple IoT devices), while smart home-10 generates low quality dataset (one IoT device). The following parameters are used for Algorithm 1 and 2: batch size K = 10 or 100, H = 1 or 3, $\alpha$ = 0.01. 

\subsection{Results}
Our goal in this work is to design a mechanism for eliciting cooperation in CDL. Cluster-based fair strategy enforces participants for cooperation in CDL; however, Theorem 1 and 2 proves that participants defect in CDL. For the unbalanced datasets, the clusters of loss values are shown in figure 3. The results show that 80\% participants collaborates with other participants, and 20\% participants learns individually by choosing $DF$ strategy in the game G. 

\begin{figure}[t]
\centering
\includegraphics[width=.5\textwidth]{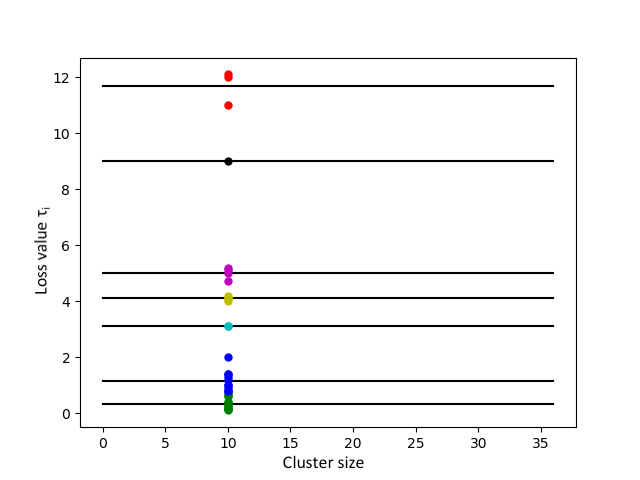}
\centering
\caption{Visualization of k-means Clustering in One Dimensional Loss Value}
\end{figure}
\section{Conclusion and Future Work}
\label{sec:conc}
In this paper, we presented a system model of CDL, and introduced the problem of strategic behavior of mobile edge devices in CDL system. We evaluated rationality of mobile edge devices in CDL using game theory model, a CDL game. We also evaluated the Nash Equilibrium (NE) strategy profile for each scenario, where the learning mobile edge devices are enforced to cooperate using our cluster-based fair strategy in CDL. We believe that this work is the first step towards a deeper understanding of the effect of non-cooperative behavior in CDL. For future work, we plan to extend the model and evaluation to determine the accuracy of ML/DL model and to train our proposed model with other IoT datasets.
{
\bibliographystyle{IEEEtran}
\bibliography{references}
}
\end{document}